\documentclass[aps,longbibliography,preprint,prb,showpacs,amsmath,superscriptaddress,onecolumn,floatfix,nobalancelastpage]{revtex4-1}

\usepackage{hyperref}
\usepackage[all]{hypcap}
\usepackage{amsmath}
\usepackage{bm}
\usepackage{graphicx}
\usepackage{grffile} % to handle dots in graphics filnames
\usepackage{color}
\usepackage{esint}
\usepackage{textpos}
\usepackage[usenames,dvipsnames]{xcolor}

\newcommand{\kk}{{\bm{k}}}
\newcommand{\im}{\textrm{Im\ }}
\newcommand{\MS}[1]{\textcolor{black}{#1}}
\newcommand{\revision}[1]{\textcolor{black}{#1}}
\newcommand{\revisiontwo}[1]{\textcolor{black}{#1}}
\newcommand{\MSNEW}[1]{\textcolor{black}{#1}}

\begin{document}
\author{Michael Sentef}
\affiliation{Stanford Institute for Materials and Energy Sciences (SIMES),
SLAC National Accelerator Laboratory, Menlo Park, CA 94025, USA}
\author{Alexander F. Kemper}
\affiliation{Lawrence Berkeley National Lab, 1 Cyclotron Road, Berkeley, CA 94720, USA}
\author{Brian Moritz}
\affiliation{Stanford Institute for Materials and Energy Sciences (SIMES),
SLAC National Accelerator Laboratory, Menlo Park, CA 94025, USA}
\affiliation{Department of Physics and Astrophysics, University of North Dakota, Grand Forks, ND 58202, USA}
\affiliation{Department of Physics, Northern Illinois University, DeKalb, IL 60115, USA}
\author{James K. Freericks}
\affiliation{Department of Physics, Georgetown University, Washington, DC 20057, USA}
\author{Zhi-Xun Shen}
\affiliation{Stanford Institute for Materials and Energy Sciences (SIMES),
SLAC National Accelerator Laboratory, Menlo Park, CA 94025, USA}
\affiliation{Departments of Physics and Applied Physics,
Stanford University, Stanford, California 94305, USA}
\affiliation{Geballe Laboratory for Advanced Materials, Stanford University, Stanford, CA 94305, USA}
\author{Thomas P. Devereaux}
\affiliation{Stanford Institute for Materials and Energy Sciences (SIMES),
SLAC National Accelerator Laboratory, Menlo Park, CA 94025, USA}
\affiliation{Geballe Laboratory for Advanced Materials, Stanford University, Stanford, CA 94305, USA}

\title{\revision{Examining electron-boson coupling using time-resolved spectroscopy}}

\begin{abstract}
Nonequilibrium pump-probe time domain spectroscopies can become an important tool to disentangle degrees of freedom whose coupling leads to broad structures in the frequency domain. Here, using the time-resolved solution of a model photoexcited electron-phonon system we show that the relaxational dynamics are directly governed by the equilibrium self-energy so that the phonon frequency sets a window for ``slow'' versus ``fast'' recovery. The overall temporal structure of this relaxation spectroscopy allows for a reliable and quantitative extraction of the electron-phonon coupling strength without requiring an effective temperature model or making strong assumptions about the underlying bare electronic band dispersion. 
\end{abstract}

\pacs{78.47.J-, 79.60.Bm, 71.38.-k}
\maketitle

\section{Introduction}
\MSNEW{
Pump-probe spectroscopies offer the exciting opportunity to perturb and measure electrons and collective modes in solids on their intrinsic ultrafast time scales. Moreover, they provide a tool to gain insights into the behavior of matter pushed out of its thermodynamic equilibrium state, thereby probing microscopic details of complex many-body systems beyond effective thermodynamic variables. The opportunity, however, comes with a challenge: When the equilibrium concepts of entropy and temperature are overcome, we are also in need to develop a language for true nonequilibrium behavior. Moreover, the value of ultrafast spectroscopies must be enhanced by the ability to relate measurements out of equilibrium to microscopically derivable quantities in an effort to understand dynamics and, ultimately, nonequilibrium itself. In this paper, we address these issues by showing a full nonequilibrium solution of a generic electron-phonon coupled model system, and by relating the temporal spectroscopic response to microscopic quantum-many body details of the system. We utilize this concept as a basis to directly access the intrinsic decay rates of quasiparticles in the time domain without assumptions about bare band quantities that are needed in the frequency domain.
}

\MSNEW{
Angle-resolved photoemission (ARPES) has been widely used to determine the low-energy electronic structure of a variety of materials.\cite{RevModPhys.75.473,Cuk2005} Many-body interaction effects are manifest via the self energy $\Sigma$ giving rise to renormalizations of the electron mass and an overall broadening of the spectral function. As an example, it was thought that ARPES data for the cuprates could thereby be studied to extract the boson coupling that is responsible for high temperature superconductivity.\cite{Bogdanov2000,Johnson2001,Lanzara2001,Kaminski2001,Dessau1991,Cuk2004,Gromko2003,Kim2003,Sato2003,Norman1997,Borisenko2006a,Borisenko2006b,Kordyuk2006,Meevasana2006,Zhou2005,Fink06,Dahm2009,Valla1999}
These investigations have provided important insights into the underlying bosons and interactions responsible for kinks in the electronic dispersion.}

\MSNEW{
However, extracting the self-energy and a meaningful electron-boson coupling constant from frequency-dependent ARPES data can be challenging. \revision{In fact there is currently no reliable way to extract a frequency-dependent electron-phonon coupling.} In addition, the interpretation of ARPES data in terms of a self-energy naturally rests upon the assumption of an underlying bare electronic band, which becomes renormalized by the coupling to bosonic modes. Similarly, the widths of momentum distribution curves (MDCs) need to be supplemented with a bare Fermi velocity in order to determine scattering rates. Estimates of the dimensionless electron-boson coupling strength $\lambda$ therefore vary strongly in the literature, from very small effective scattering rates\cite{Reber12} to extremely strong coupling\cite{Fink06}.}

As an alternative, a time domain approach using the relaxation of excited electrons back to equilibrium as a direct technique to measure electron-phonon coupling was proposed by Allen.\cite{Allen1987}
The basic quantities for this method are the deviations from the initial equilibrium values of the electron and phonon occupations in momentum space. A semi-classical Boltzmann approximation that averages over momenta and energy to extract a single energy relaxation time scale was used to extract estimates for electron-phonon couplings from pump-probe photoemission data in simple elemental metals,\cite{PhysRevLett.64.2172,0953-8984-19-8-083201} but it was also applied to analyze pump-probe data for high-temperature superconductors.\cite{Perfetti2007,Cortes2011,Graf2011,Rettig2012,Avigo13} 

However, in correlated materials, such as transition metal oxides like the cuprates, or in multiband systems with multiple disconnected Fermi surface pieces, such as the iron pnictides, the momentum and energy dependences must be resolved, since \MS{they are crucial ingredients} for these materials' interesting properties.\cite{Kemper2011,Devereaux2004} In addition, a full quantum mechanical formulation is required to treat arbitrarily strong fields and strong energy and momentum dependent scattering, particularly in materials where the balance between various competing phases may be tipped with the application of strong electric fields. Examples include possible light-induced superconductivity in a stripe-ordered cuprate,\cite{Fausti14012011} a dramatically enhanced conductivity in an oxide heterostructure,\cite{Caviglia2012} coherent phonon oscillations serving as a marker for a photoinduced phase transition in VO$_2$,\cite{Wall2012}, the theoretical proposal of dynamical band flipping,\cite{PhysRevLett.106.236401} and pump-probe photoemission providing a novel classification scheme for charge-density wave insulators in the time domain.\cite{Hellmann2012} 

\MSNEW{
In this paper, we combine the strength of the time domain approach with the power of ARPES self-energy microscopics.} We present a microscopic quantum nonequilibrium approach with full momentum and energy resolution that can be applied when effective temperature phenomenology and the semiclassical Boltzmann equation fail, and that can be used in conjunction with standard many-body and material specific computational approaches. Using our model system, we show how the time ($t$), momentum ($\kk$), and energy ($\omega$) resolution provided by state-of-the-art tr-ARPES \cite{Fann1992,Hentschel01,Schmitt2008,Perfetti06,Perfetti08,Perfetti2007,Dakovski10,Cortes2011,Graf2011,Rettig2012,Sobota2012,McIver2012,Heyl12,Hellmann12,Rohde13} experiments in conjunction with the dependence of extracted decay times on the initial equilibrium sample temperature provides a direct method to measure the dominant phonon energy and electron-phonon coupling.
Hence, we show that relaxational spectroscopy in the time domain allows the possibility to extract detailed information about the system in equilibrium and lends itself to a simple interpretation that does not involve any assumptions about the bare electronic band dispersion and goes beyond the traditional two-temperature model phenomenology.

\MSNEW{
The paper is organized as follows. In Section \ref{Methods} we outline the electron-phonon coupled model system and the calculation of double-time Green functions  on the Keldysh contour and the time-resolved photoemission response. In Section \ref{Results} we present our results for the photoexcited electronic structure (\ref{Structure}) and the extraction of the quasiparticle self-energy and the implied electron-phonon coupling from the temporal relaxation of the electrons (\ref{Extracting}). Section \ref{Discussion} contains a discussion of the results and their implications. Additional considerations regarding the tradeoff between time and energy resolution, the breakdown of an effective temperature description, the connection between relaxation rates and the self-energy, and the additional effect of including electron-electron scattering are provided in a Supplementary Material.
}

\MSNEW{
\section{Model and Method}
}
\label{Methods}

\subsection{Introductory remarks}
For the formulation of the nonequilibrium problem, we employ the well-established Keldysh Green function technique. The single-particle properties of the electrons are described by the double time Green function $G(t,t')$ on the complex Keldysh contour. These encode the electrons' initial equilibrium state, their real time evolution during and after a perturbation via a pump laser pulse, and their memory of the initial state. The double-time Green functions allow us to extract the relevant information (occupation of states, tr-ARPES response) about the excited electron system in a microscopic way. 

The pump laser field of arbitrary temporal shape is included as a space-independent and time-dependent vector potential $\bm{A}(t)$ in a nonperturbative way via Peierls substitution.\cite{Peierls1933,Jauho} This approach has been formulated earlier by Jauho and Wilkins\cite{Jauho} which we extend numerically to longer time scales and arbitrary field strengths.  The coupling of the electrons to a nondispersive optical phonon mode is modeled by the well-established and controlled diagrammatic Migdal approximation.\cite{Migdal} The resulting integro-differential equations of motion for the Green functions\cite{Eckstein10} are solved using massively parallel numerical integration. The code is parallelized easily since the equations for different momentum points decouple in the discretized Brillouin zone. An additional Fourier transform of the lesser component of the resulting interacting Green function is performed to extract the tr-ARPES spectra.\cite{Freericks09TRARPES} In this step, a shape function is introduced to take the probe pulse width into account. Additionally, the momenta need to be time-shifted depending on the pump field to ensure gauge invariance.\cite{Turkowski07BOOK}

\subsection{Model}
We study the Holstein Hamiltonian\cite{Holstein}
\begin{align}
H =& \sum_{\bm{k}} \epsilon(\bm{k}) c_{\bm{k}}^{\dagger}c_{\bm{k}}^{} + \Omega \sum_i b_i^{\dagger}b_i^{} - g \sum_i c_i^{\dagger} c_i^{} (b_i^{} + b_i^{\dagger}),
\end{align}
where the individual terms are the electronic kinetic energy with dispersion $\epsilon(\kk)$, the phonon energy for dispersionless Einstein phonons of frequency $\Omega$, and the electron-phonon interaction, respectively. Here $c_{\kk(i)}^{(\dagger)}$ is the electronic annihilation (creation) operator in momentum space (real space), $b_i^{(\dagger)}$ is the bosonic phonon annihilation (creation) operator on lattice site $i$, and $g$ is the momentum-independent electron-phonon coupling strength. 

The starting bare nonequilibrium Green function\cite{Turkowski07BOOK}
\begin{align}
G^0_\kk(t,t') =& i \left[ n_F(\epsilon(\kk)) - \theta_c(t,t') \right]
\nonumber\\
&\times
\exp\left[ -i \int_{t'}^t d\bar t\ 
\epsilon\left(\kk -\bm{A}(\bar t)\right) \right]
\end{align}
where $n_F(\omega)$ $=$ $1/(1+\exp(\omega/T))$ is the Fermi function at temperature $T$, $t$ and $t'$ are times on the Keldysh contour $\mathcal{C}$, and $\theta_c(t,t')$ is the Heaviside function on the contour, which equals 1 if $t$ is later on the contour than $t'$ and 0 otherwise. $\epsilon(\kk)$ is the single-particle energy dispersion for a square lattice with nearest ($V_{nn}$) and next nearest neighbor ($V_{nnn}$) hopping,
\begin{align}
\epsilon(\kk) = -2 V_{nn} \left( \cos k_x + \cos k_y \right) + 4 V_{nnn} \cos k_x \cos k_y - \mu,
\end{align}
measured with respect to the chemical potential $\mu$. Here, we use the
convention that $\hbar=c=e=1$, and we work in the Hamiltonian gauge, i.e. the scalar potential is set to zero. 

We incorporate the electron-phonon interactions in the Migdal limit, which is appropriate for weak coupling. In this case the electronic self-energy due to electron-phonon coupling is
\begin{align}
\Sigma(t,t') = i g^2 \sum_\kk D^0(t,t') G^0_{\kk}(t,t').
\end{align}
The bare phonon Green's function $D^0(t,t')$ is\cite{Mahan}
\begin{align}
D^0(t,t') =& -i \left[ n_B(\Omega) + 1 - \theta_c(t,t') \right] \exp\left( i\Omega(t-t') \right) \nonumber \\
&-i\left[n_B(\Omega) + \theta_c(t,t')\right] \exp\left( -i\Omega(t-t') \right )
\end{align}
where $n_B(\Omega)$ $=$ $1/(\exp(\Omega/T)-1)$ is the Bose function. 

\subsection{Method}
With the self-energy above, we solve the Dyson equation for the interacting Green function $G$,
\begin{align}
G_\kk(t,t') = G^0_\kk(t,t') + \int_{\mathcal{C}} dt_1 dt_2 G^0_\kk(t,t_1) \Sigma(t_1,t_2) G_\kk(t_2,t').
\end{align}
This can be done by casting the Dyson equation as a matrix equation.  As an alternative, the Dyson equation can be rewritten by expanding the integral with Langreth rules, which gives equations of motion for the Matsubara, retarded, real-imaginary,
and lesser Green's functions.\cite{Wagner91,Eckstein10} This leads to a set of Volterra
integro-differential equations that can be solved via standard numerical integration. For the results presented here, we have used the Volterra scheme. Its main advantages are that the equations are manifestly causal, and the required memory is reduced by the use of symmetries.

The pulse that is of direct interest to pump-probe experiments is, by nature, a propagating light
pulse; this implies an oscillating field without a zero-frequency component.
We model the pump via an oscillating vector potential along the $(11)$ or Brillouin zone diagonal direction with a Gaussian
profile,
\begin{align}
\bm{A}(t) = (\bm{\hat x + \hat y})\frac{\mathrm{{E}_{max}}}{\omega_p} \sin(\omega_p t) \exp\big(-\frac{(t-t_0)^2}{{2}\sigma^2}\big).
\end{align}
The tr-ARPES intensity for a probe pulse of width $\sigma_{pr}$ is computed from\cite{Freericks09TRARPES}
\begin{align}
I(\kk,\omega,t_0) =& \im \frac{1}{2\pi\sigma_{pr}^2} \int dt \int dt' \tilde{G}^<_\kk(t,t')
\nonumber\\
& \times e^{-(t-t_0)^2/2\sigma_{pr}^2} e^{-(t'-t_0)^2/2\sigma_{pr}^2} e^{i\omega(t-t')},
\end{align}
where the integral is along real times, and each component $\kk_{i}$ of the momentum argument of $\tilde{G}^<_\kk(t,t')$ is shifted ($\kk_{i}$ $\rightarrow$ $\kk_{i}-\frac{1}{t'-t}\int_t^{t'} d\bar{t} A_i(\bar{t})$, where $A_i$ is the corresponding $i$-th component of the vector potential) to restore gauge invariance.\cite{Turkowski07BOOK} In the Supplementary Material, we illustrate the importance of the temporal probe pulse width $\sigma_{pr}$ for the effective energy resolution of spectral features.

\subsection{Setup and model parameters}
We consider 2D tight-binding electrons in a partially filled metallic band with parameters $V_{nn}=0.25$ eV, $V_{nnn}=0.075$ eV, $\mu=-0.255$ eV, chosen to approximate a Fermi surface as seen in the cuprates at high dopings where the pseudogap is negligible, for example.. The dispersion less Holstein phonons have a frequency $\Omega$ $=$ 0.1 eV and electron-phonon coupling $g$ $=$ $\sqrt{0.02}$ eV as representative values for optical phonons in the cuprates. In equilibrium, the phonon frequency sets the relevant energy scale at which a kink at $\pm \Omega$ in the band structure occurs\cite{Ashcroft}, as shown in Figure 1a-c. The dimensionless coupling constant $\lambda$ is a measure of the strength of lattice distortion relative to the kinetic energy of the electrons, and is defined via the slope of the real part of the self-energy that determines the electronic mass renormalization. Our choice of parameters corresponds to a value of $\lambda$ $=$ 0.42, which is sufficiently large to produce a visible kink in the dispersion for our balance of energy and time resolution, but still small enough to justify the use of the Migdal approximation for the electron-phonon coupling. 

For the electric pump and probe fields, we choose parameters that allow us to compute the full time evolution on the Keldysh contour by the solution of the integro-differential equations with a reasonable computational effort. Specifically we choose a Gaussian field envelope of width $\sigma$ $=$ 4 femtoseconds (fs) and a center frequency of 0.5 eV, which is smaller than UV pump pulses around 1.5 eV, which are currently typically used in tr-ARPES experiments. The smaller frequency affects the transient behavior quantitatively. However, it allows us to use a coarser time step discretization and therefore to observe the relaxation of the excited electrons for a longer time, while it does not affect our conclusions about relaxation effects. The delay time is defined with respect to the center of the Gaussian pulse envelope. The electric field has a peak strength $E_{\text{max}}$ $=$ 0.4 V/$a_0$. While this field strength is at the high end of what can be currently achieved experimentally, this choice is motivated by the fact that it leads to relatively strong, clearly visible spectral redistributions whose temporal decay can be followed easily. Again, the details of the pump do affect the results, but not the conclusions. This is explicitly demonstrated in the Supplement by analyzing the relaxation times for different maximal pump field strengths.

The subsequent tr-ARPES probe pulse has a Gaussian envelope width of $\sigma_{pr}$ $=$ 16.5 fs. This choice is motivated by the maximum contour length that can be reached with reasonable computational effort, since the earliest and latest tr-ARPES observation delay times for which one can extract spectra have to fulfill a 3 $\sigma_{pr}$ minimum distance to the actual earliest and latest times on the contour. In the Supplement, we also show the equilibrium tr-ARPES spectra for smaller $\sigma_{pr}$, which enables a better time resolution but necessarily leads to worse energy resolution.

In the simulations, energy scales are measured in units of eV, which implies a natural time scale of h/eV $\approx$ 0.658 fs. While the scales are fixed here for clarity, one should keep in mind that a simple rescaling can be applied, e.g., doubling all the time scales while dividing the energy scales in half preserves all the relations.

\section{Results}
\label{Results}
\subsection{Photoexcited electronic structure: Energy and temperature dependent relaxation}
\label{Structure}
In Figure 1, we show the energy- and momentum-resolved tr-ARPES spectra along a nodal cut for three delay times before (1a), during (1b), and after the pump pulse (1c). The pump pulse drives electrons from below the Fermi level $E_F$ to empty states above $E_F$ (Figure 1b) as the system absorbs energy leaving the electrons in a photoexcited state. The excess energy of the electrons is then transferred into the bath of phonons, and the electrons relax back towards equilibrium (Figure 1c). Figure 1d provides a comparison of the $\kk$-resolved energy distribution curves (EDC) before and 89 fs after the pump. Clearly, the phonon window $\mathcal{W}$ $=$ $[-\Omega,\Omega]$ around the Fermi level sets an energy scale for persistent spectral changes: while electronic quasiparticles outside this window have already relaxed back to their equilibrium EDC, the quasiparticles inside remain excited at the particular time delay shown here. This phonon window effect is highlighted in the momentum-integrated pump-induced spectral changes in Figure 1e: photoexcited electrons above $E_F$ and holes below $E_F$ only persist inside the window set by the phonon frequency $\Omega$. 

In equilibrium at zero temperature, kinematics prevents electron relaxation unless the quasiparticles have sufficient energy to emit phonons, which is why this window is expected. In Figure 1f, we show that raising the initial equilibrium temperature leads to smaller spectral changes compared to Figure 1d at the same time delay. Correspondingly, the $\kk$-integrated spectral changes shown in Figure 1g are significantly smaller than the ones in 1e (shown on the same vertical scale), although the phonon window effect is still visible.

\MSNEW{We note here that this result is quite reminiscent of electronic relaxation measured in graphene \cite{Winnerl11,Gilbertson12,Winnerl13, Johannsen13,Gierz13} as well as in an optimally doped cuprate superconductor in the normal and superconducting states.\revision{\cite{Graf2011,Smallwood2012}} The phonon window effect provides guidance towards a quantitative way of understanding binding energy resolved pump-probe decay rates in terms of the equilibrium self-energy. Before presenting the details of this analysis, we first show that a more traditional way of interpreting pump-probe spectroscopies, the ``hot electron'' model, is neither necessarily appropriate for describing the nonequilibrium transient state nor a prerequisite for our method to work.} The hot electron model assumes that the laser pump initially raises the temperature of the electrons in a hot quasithermal state practically decoupled from the lattice. The quasithermalization of electrons is facilitated on short time scales by electron-electron scattering, leading to distinct electronic and lattice ``temperatures'' in two-temperature models.\cite{Anisimov1974,Allen1987} The electrons then transfer their excess energy to the lattice until the electronic and lattice subsystems equilibrate. 

For our model system, the hot electron picture does not provide a valid description of the photoexcited electrons, which is demonstrated in Figure 2, \MS{because we do not have any electron-electron scattering to allow for the fast relaxation of the electrons into a thermal distribution.} In Figure 2a and 2b, we \MS{demonstrate the influence of the electric field strength by showing} the momentum-resolved tr-ARPES spectrum along the diagonal momentum cut at a short time delay (1 fs) for two different pump pulses. Figure 2a is for the same maximum field strength $E_{\text{max}}$ $=$ 0.4 V/$a_0$ that was used in Figure 1; Figure 2b is for a ten times stronger field. The weaker pump mainly excites electrons within the bare energy band, but the stronger pump clearly disrupts the band structure and induces ladder-like structures which are reminiscent of Wannier-Stark ladders in strong dc electric fields.\cite{Mendez1993,Moritz2010} 

The breakdown of the ``hot electron'' picture becomes evident when we integrate the tr-ARPES spectra in the momentum window around the Fermi momentum between $\bm{k}_1$ and $\bm{k}_7$ (see Figure 1). The energy distribution curves in Figure 2c are not simply featureless decaying exponentials, as they would be if the ``hot electron'' model was valid, but show distinct higher energy peaks related to the pump characteristics. Although it is possible to extract an effective temperature from the linear slope of the logarithmic intensities in a very small energy window ($< 0.1$ eV) near the Fermi level (see Figure S3 in Supplement), this ``temperature'' does not provide a meaningful description of the photoexcited system. In particular, it fails to account for the actual excess energy in the electronic subsystem due to the spectral features at higher energies. In fact, once this high spectral weight has dissipated, a fit to an effective Fermi distribution shows that the increase in the effective temperature is very small, much smaller than that obtained from forcing a fit a shorter time delays (see Supplement).

These nonthermal features are even more pronounced for the strong pump. This observation clearly shows that photoexcitation is not necessarily the same as heating, and that a full nonequilibrium modeling beyond effective temperature phenomenology is often essential for correctly describing pump-probe spectroscopies. \MS{Since we include only electron-phonon scattering in our simulation and neglect electron-electron scattering, the discrepancy between quasithermal behavior and the observed transient spectra is not too surprising. The inclusion of electron-electron scattering leads to a closer resemblance of spectra with two-temperature phenomenology (see additional calculations including electron-electron scattering in the Supplement), but this is not guaranteed to be the case. Moreover, even if it is the case, the single relaxation rate from two-temperature fits is not simply connected to energy dependent population decay rates, as we show in the Supplement. Therefore, since tr-ARPES experiments probe the full energy- and momentum-resolved population dynamics, the data analysis should go beyond effective temperature decay rates.}
 
\subsection{Extracting equilibrium system properties by pump-probe photoemission}
\label{Extracting}
We now focus on a quantitative analysis of energy-dependent relaxation rates for the weaker pump and the higher initial temperature $T$ $=$ 290 K. Figure 3a shows the pump-induced spectral changes relative to the initial state, integrated along momenta on the Brillouin zone diagonal, in a continuous map as a function of energy and delay time. Here, the phonon window effect becomes even more obvious: persistent excitations of electrons and holes are only observed inside $\mathcal{W}$, whereas photoexcited electrons outside $\mathcal{W}$ decay on a much faster time scale. The dashed lines mark the phonon window $\mathcal{W}$, which clearly sets the relevant energy scale for slow versus fast relaxation. In Figure 3b, we present cuts for selected energies of the same data as in Figure 3a: Both electrons and holes inside $\mathcal{W}$ decay on a significantly longer time scale than their counterparts outside $\mathcal{W}$. The slight asymmetry between holes and electrons (negative and positive binding energies) is related to asymmetry in the underlying band structure, made clear in the discussion which follows. 

We fit decaying exponentials (solid lines in Figure 3b) to these curves starting at a delay time of 40 fs, chosen to be sufficiently large such that pump and probe pulses do not significantly overlap. From these fits we extract the tr-ARPES relaxation rates shown in Figure 4a on a logarithmic scale. The relaxation rates show a steep decline inside the phonon window. This characteristic behavior indeed is reminiscent of the \emph{equilibrium} scattering rates (self-energy) due to coupling between electrons and an optical phonon branch, reflected in a suppression of the imaginary part of the equilibrium self-energy in the phonon window. To make this connection between tr-ARPES relaxation rates and the equilibrium self-energy apparent, we plot $-2\;\text{Im}\; \Sigma_{\text{aver}}$ in Figure 4a. For comparison we account for energy resolution due to the probe pulse width by an additional averaging of relaxation times appropriate for the pump-probe setup (see Supplement). 

The good agreement between the tr-ARPES relaxation rates and the equilibrium self-energy is the central result of our paper. It indicates that the return to equilibrium of an electron excited into a state of energy $\omega$ and momentum $\bm{k}$ is solely determined by the equilibrium quasiparticle self-energy $\Sigma(\bm{k},\omega)$. This result suggests that relaxational pump-probe photoemission spectroscopy can be used as an effective tool for accurately measuring equilibrium system properties in a direct way in the time domain without resorting to any assumptions about the underlying bare band structure. Traditionally, equilibrium ARPES experiments are used to extract an estimate for electronic lifetimes through the linewidths of momentum distribution curves (MDCs). For optimal energy and momentum resolution and only a single source of decoherence, the full width at half maximum (FWHM) is related to the relaxation rate at the corresponding energy via $\tau^{-1}(\omega)$ $=$ $\text{FWHM}(\omega) v_F$, where $v_F$ is the bare Fermi velocity along the corresponding momentum cut and the equation is valid near the Fermi level for a linear band dispersion. 
In order to compare relaxation rates extracted from MDC widths directly for the same energy resolution as the tr-ARPES decay rates, we extract the MDC widths from a tr-ARPES snapshot using the same probe width ($\sigma_{pr}$ $=$ 16.5 fs) as for the pump-probe simulation as well as for a continuous probe beam as in a true equilibrium ARPES experiment. The continuous probe equilibrium MDC widths show good agreement with the tr-ARPES decay rates, but the MDC linewidths for the pump-probe setup with restricted energy resolution imply larger rates inside $\mathcal{W}$ (see Figure 4a). This can be understood from the fact that when the relaxation rates and therefore the MDC widths vary strongly with binding energy, the energy resolution has a crucial effect on the effective imaginary part of the self-energy extracted from MDC widths.

%CHANGE
It might seem counterintuitive that a system out of equilibrium relaxes at the equilibrium scattering time scale, determined by the imaginary part of the self-energy, but this result should hold in nearly all systems at long times provided that the system relaxes back to an equilibrium state.
\revisiontwo{
This has been shown for single-time kinetic equations using a relaxation time ansatz\cite{Mahan87,Allen1987} (see also the appendix of Ref.~\onlinecite{Kemper13}). However, by construction, a single time formalism is incapable of addressing the momentum \emph{and energy} dependent relaxation rates measured with pump-probe photoemission spectroscopy. As we have shown, the energy dependence is crucial to extract information about dominant bosonic mode couplings which appear as the phonon window effect. Both the computation of the tr-ARPES response and the connection of its relaxation rates with the dynamical equilibrium self-energy require the full double time Green function formalism used in this work.
}

\revisiontwo{
For the full double time Green functions, one can show that nonequilibrium relaxation is governed by the equilibrium self-energy along the following lines:}
For small deviations from equilibrium the change of the spectrum is second order in the deviation from equilibrium, while the change of the occupation is first order, primarily due to the change of the distribution function from the equilibrium Fermi-Dirac distribution.\cite{Mahan87,Freericks06} \revisiontwo{This result arises from a linearization of the quantum Boltzmann equation for small perturbations from equilibrium and explains why linear response coefficients can be calculated from equilibrium response functions.} Hence, the Green functions should be the same as in equilibrium, except for a small change of the distribution function. If the self-energy of the system in equilibrium does not depend too strongly on temperature, then it is weakly dependent on the distribution function, and as a result, we will see the nonequilibrium system relax back to equilibrium at the longer time scales via the equilibrium relaxation rates. These conditions are usually fulfilled for sufficiently high temperature. Care has to be taken when the temperature becomes so low compared to the lowest relevant bosonic mode energy that relaxation via the equilibrium self-energy is completely blocked at low binding energies. In this case, relaxation pathways of higher order in the electron-boson coupling become relevant for the long-time relaxation, corresponding to effects of higher order in the deviation from the equilibrium distribution. 

We now show how more detailed knowledge about the electron-phonon scattering processes allows us to quantitatively extract the electron-phonon mode coupling strength from a comparison of the tr-ARPES relaxation rates $\Gamma_{\text{tr}}(\omega)$ and the self-energy $\Sigma(\omega)$. To this end, we treat our tr-ARPES relaxation rates as ``experimental data'' and assume that we know the line shape of $\text{Im}\;\Sigma(\omega)$. Within the Migdal approximation to the Holstein model, the relaxation rate is proportional to $g^2$ and of the form $\Gamma(\omega)$ $=$ $g^2$ $\phi(\omega)$ (see Supplement). The function $\phi(\omega)$ reflects electron-phonon scattering processes and takes into account the available scattering phase space related to the electronic band structure via the local electronic density of states $N(\omega)$. This is the reason why an asymmetry in the band structure results in an electron-hole asymmetry in the relaxation rates.

The measured tr-ARPES relaxation rates $\Gamma_{\text{tr}}$ are naturally convolved with the probe pulse shape and therefore include an additional energy averaging, since there is a tradeoff between time resolution and energy resolution. Therefore we apply an energy averaging procedure to $\phi(\omega)$ (see Supplement), which yields $\phi_{\text{aver}}$. The electron-phonon coupling strength is then extracted from $g^2$ $=$ $\Gamma_{\text{tr}}(\omega)/\phi_{\text{aver}}(\omega)$. Figure 4b shows this ratio as a function of energy, which serves as an \emph{a posteriori} estimate of $g^2$ given the numerical simulation data. The extracted mean value including an error estimate from the statistical standard deviation is $g^2$ $=$ 0.021 $\pm$ 0.004 eV$^2$, to be compared with the exact input value $g^2$ $=$ 0.02 eV$^2$. This agreement highlights the fact that a nonequilibrium measurement can serve as an accurate tool for extracting equilibrium  system properties, a feat which may be difficult to accomplish by pure equilibrium measurements.

\section{Discussion}
\label{Discussion}
The central result of this paper is that equilibrium ideas can be used to understand the nonequilibrium dynamics of a pumped system. This is facilitated by the connection between relaxation rates for the decay of photoexcited electrons and the equilibrium electron-phonon scattering rates. Importantly, both the dependence of the rates on the binding energy and on the initial equilibrium temperature affect the result, but not the conclusion. We have shown that a simplified modeling in terms of effective temperatures does not provide an adequate description of photoexcited systems with full nonequilibrium quantum dynamics. \MS{This may partly be due to the neglect of electron-electron scattering in our study. However, even if the effective temperature model fits work, it boils down rich population dynamics into a single temperature decay rate. Here we have presented a more appropriate way of analyzing the full wealth of momentum- and energy-dependent population dynamics measured by time- and angle-resolved photoemission spectroscopy.} 

The relation between the equilibrium self-energy and the nonequilibrium relaxation makes pump-probe spectroscopies a convenient tool to disentangle equilibrium decoherence pathways. As a consequence, the strength of electron-phonon couplings can be extracted from pump-probe relaxation times with a high level of accuracy. \MS{The presence of electron-electron interactions is expected to lead to richer short-time transient behavior, but supports this conclusion, provided that the electronic system relaxes back to an equilibrium state at long times by dissipating the pump energy into the lattice. This is shown in the Supplement by adding electron-electron interactions as an additional scattering channel.} For dispersive boson modes like acoustic phonons, the window effect is not as pronounced as for the Holstein mode discussed in this paper, but the same simulation technique and analysis can be applied. The same is true for multiple bosonic modes coupled to the electrons, each of which will show up as a characteristic change of the decay rate around the respective mode energy. Finally, our approach also can be applied to general electron-boson interactions, for instance coupling to magnons. It could therefore potentially be used to better characterize the bosonic couplings in high-temperature superconductors.

As a final remark, the recent work by Kemper \emph{et al.~}\cite{Kemper13} provides a complementary point of view by demonstrating that time domain Compton scattering measures characteristic oscillations of the time-dependent momentum distribution. These oscillations correspond to energy shifts due to bosonic mode couplings and thereby provide information on the real part of the self-energy, which is related through analyticity by the Kramers-Kronig relations to the imaginary part. Since the imaginary part is measured by time-resolved photoemission as discussed in the present work, the two works together show the power of time-resolved spectroscopies, which demonstrably bear the potential to pinpoint the full nontrivial many-body dynamics of complex quantum systems beyond equilibrium probes. 

\section{Movies}
Two movies are attached to this manuscript. They show the full time-resolved ARPES signal for the weak and strong pumps corresponding to the snapshots shown in Figure 2a and 2b. The unit for the time label shown in the movies is 1 time unit $\widehat{=}$ h/eV $\approx$ 0.658 fs.

\section*{ACKNOWLEDGMENTS}
This work was supported by the Department of Energy, Office of Basic Energy Sciences, Division of Materials Sciences and Engineering (DMSE) under Contract Nos. DE-AC02-76SF00515 (Stanford/SIMES), DE-FG02-08ER46542 (Georgetown), and DE-SC0007091 (for the collaboration). Computational resources were provided by the National Energy Research Scientific Computing Center supported by the Department of Energy, Office of Science, under Contract No. DE- AC02-05CH11231. J.K.F. was also supported by the McDevitt bequest at Georgetown.

\bibliography{bib}{}

%%%%%%%%%%%%%%%%%%%%%%%%%%%%%%%%%%%%%%%%%%%%%%%%%%%%%%%
%%%%% FIGURES     OF        MAIN          TEXT %%%%%%%%%%%%%%%%%%%%%%%%%%%%%%%
%%%%%%%%%%%%%%%%%%%%%%%%%%%%%%%%%%%%%%%%%%%%%%%%%%%%%%%

%
\newpage
\begin{figure*}[h!t]
	\includegraphics[width=0.65\textwidth]{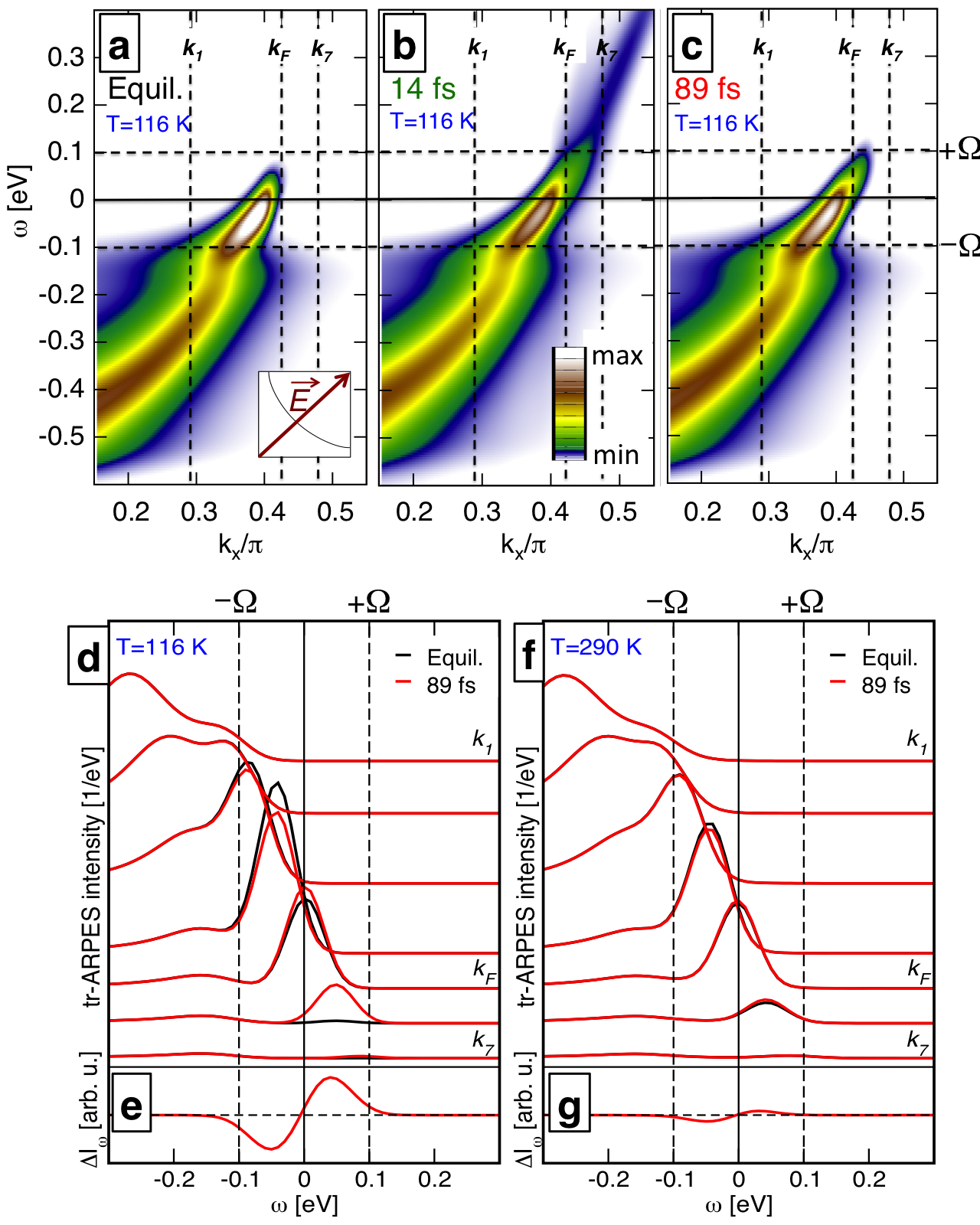}
        \caption{{\bf $\bm{k}$-resolved photoexcited electronic structure.} (a) Equilibrium tr-ARPES intensity at temperature $T$ $=$ 116 K for a cut along the Brillouin zone diagonal. The pump field polarization is aligned along this direction (inset). Horizontal dashed lines mark the position of ``kinks'' in the electronic dispersion at $\pm \Omega$. (b) Map of the tr-ARPES intensity shortly (14 fs) after the pump pulse and (c) at a later delay time $t$ $=$ 89 fs. The system has partially relaxed toward equilibrium outside the energy range $\pm \Omega$, but quasiparticles remain excited within this range. (d) Energy distribution curves (EDCs) corresponding to the data shown in (a, equilibrium) and (c, 89 fs) for momenta $\bm{k}_1$ to $\bm{k}_7$ as indicated in panels (a) and (c). For clarity EDCs are vertically shifted. (e) Intensity difference $\Delta I_{\omega}$ between photoexcited (89 fs) and equilibrium tr-ARPES data. (f),(g) EDCs and spectral changes (same as in (d),(e)) for a higher initial equilibrium temperature $T$ $=$ 290 K, showing that the time for recovery shortens for increased temperature.}
        \label{fig2}
\end{figure*}
\newpage
\begin{figure}[h!t]
        \includegraphics[width=0.85\columnwidth]{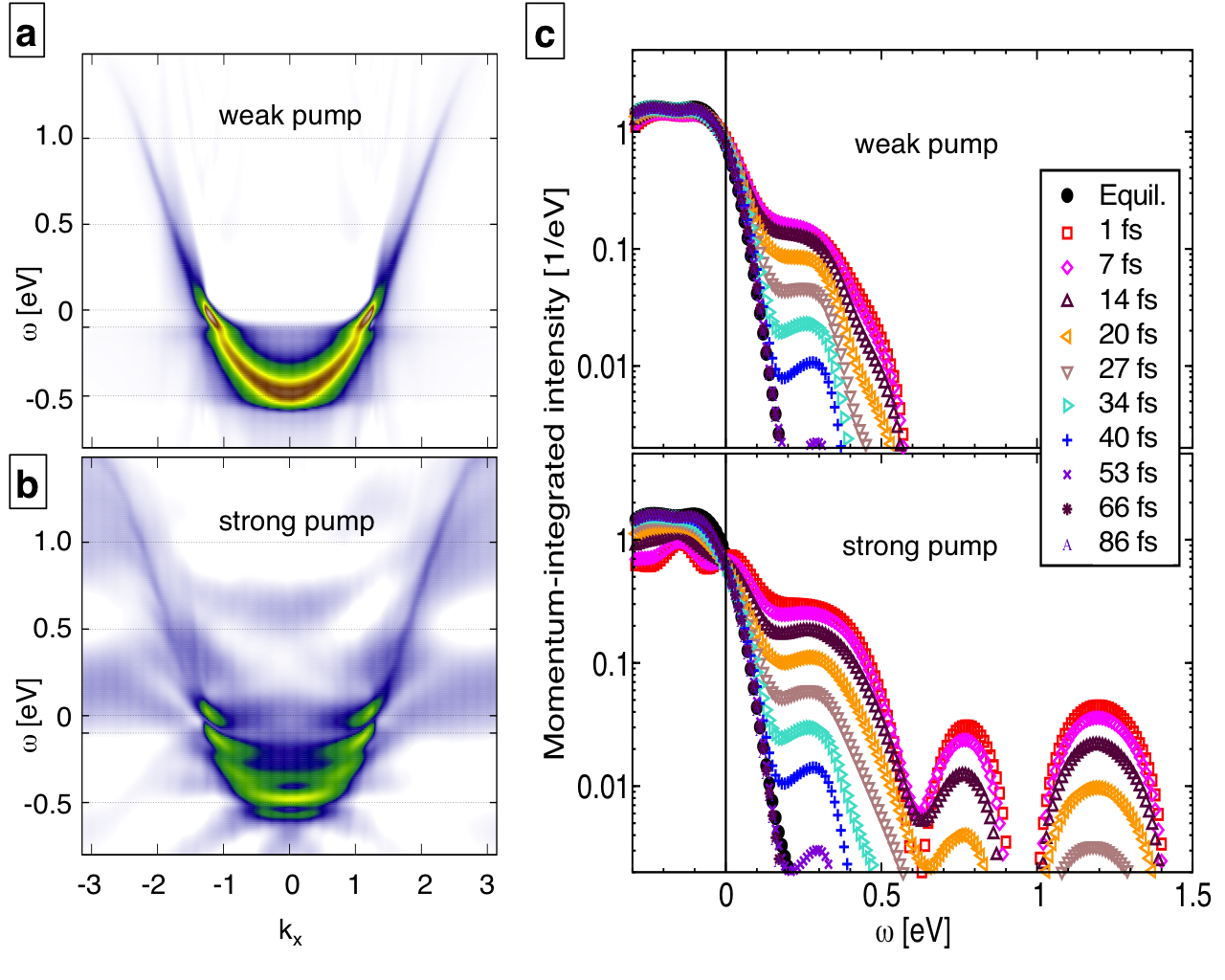}
        \caption{{\bf Weak versus strong pump field.} {\bf a,} tr-ARPES intensity map (delay time of 1 fs) for a weak pump field (same as in Figure 1) with maximum field strength $E_{\text{max}}$ $=$ 0.4 $\text{V}/a_0$.
        {\bf b,} Ten times stronger pump field ($E_{\text{max}}$ $=$ 4 $\text{V}/a_0$). For a lattice constant $a_0$ $=$ 3.4 $\text{\AA}$, these peak field strengths are $\approx$ 12 MV/cm (``weak pump'') and 118 MV/cm (``strong pump''), respectively.
        {\bf c,} Energy distribution curves integrated over momenta along the Brillouin zone diagonal (same momentum window $\bm{k}_1$ to $\bm{k}_7$ close to $\bm{k}_F$ as indicated in Figure 1) for various pump-probe delay times. 
        }
        \label{figsupp3}
\end{figure}
\newpage
\begin{figure*}[h!t]
	\includegraphics[width=\textwidth]{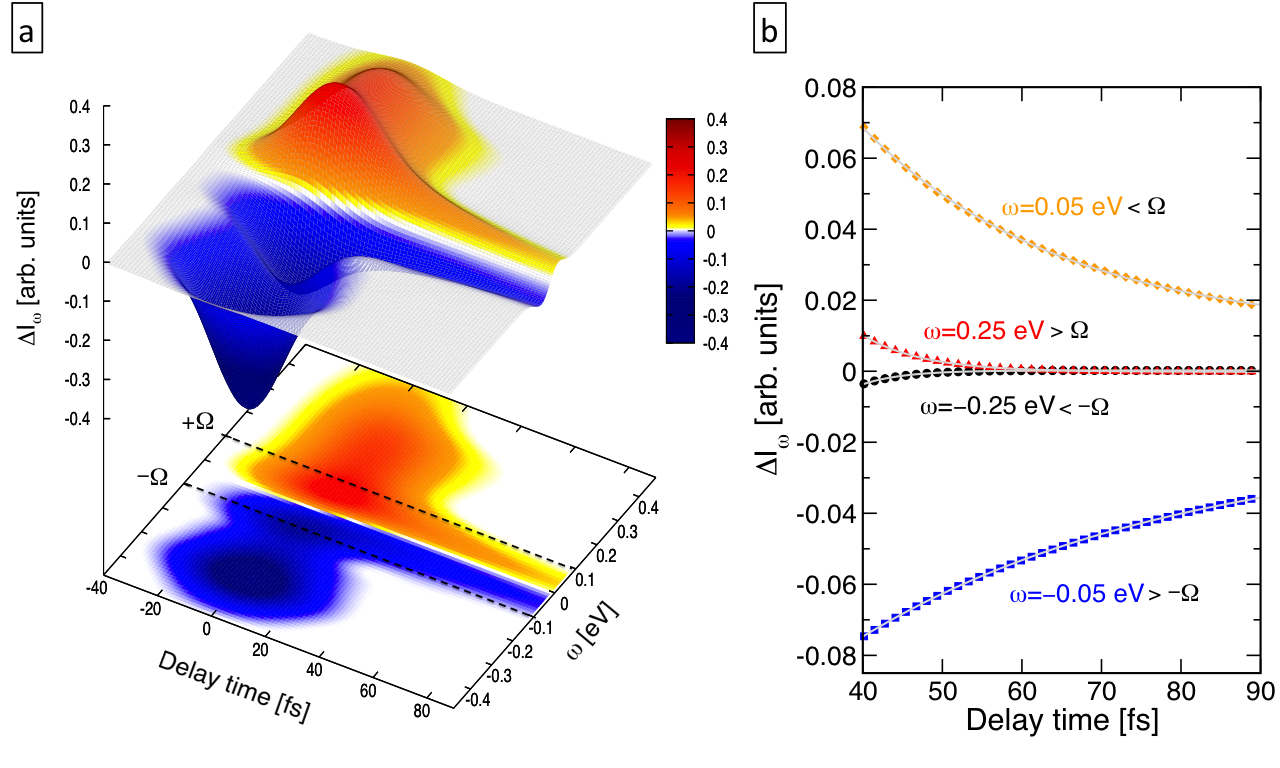}
        \caption{{\bf Transient relaxation of the pump-induced spectral intensity changes for equilibrium temperature $T$ $=$ 290 K.}
        (a) Spectral changes as a function of pump-probe delay time integrated over momenta along the whole Brillouin zone diagonal. 
        (b) Same data as in {\bf a} for selected energies inside (0.05 eV) and outside (0.25 eV) the phonon window for positive (electrons) and negative (holes) binding energies.
        }
        \label{fig3}
\end{figure*}
\newpage
\begin{figure*}[h!t]
        \includegraphics[width=\textwidth]{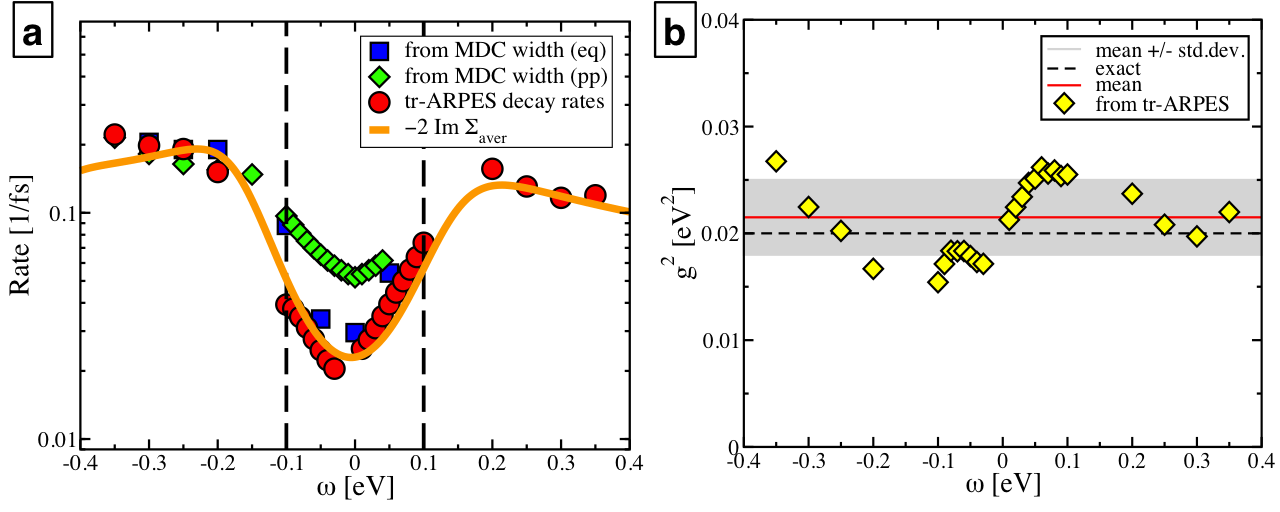}
        \caption{{\bf Energy dependent relaxation due to electron-phonon scattering.}
        (a) Relaxation rates from tr-ARPES (red circles) compared with equilibrium MDC widths for a continuous probe beam as in an equilibrium measurement (``eq'', blue squares) and pump-probe setup (``pp'', $\sigma_{pr}$ $=$ 16.5 fs probe pulse, green diamonds). The orange line shows the corresponding equilibrium electron-phonon scattering rate taking into account energy resolution (see Supplement). 
        (b) Extraction of the electron-phonon coupling strength from the ratio of tr-ARPES decay rates and equilibrium self-energy (see Supplement). The red solid line shows the mean value of the data points. The black dashed line indicates the exact input value of $g^2$. The grey shaded area shows the confidence interval estimated from the statistical standard deviation of the data.
        }
        \label{fig4}
\end{figure*}
%
%%%%%%%%%%%%%%%%%%%%%%%%%%%%%%%%%%%%%%%%%%%%%%%%%%%%%%%
%%%%% END FIGURES     OF        MAIN          TEXT %%%%%%%%%%%%%%%%%%%%%%%%%%%%%
%%%%%%%%%%%%%%%%%%%%%%%%%%%%%%%%%%%%%%%%%%%%%%%%%%%%%%%

\end{document}

\author{Michael Sentef}
%\affiliation{Stanford Institute for Materials and Energy Science (SIMES),
%SLAC National Accelerator Laboratory, Menlo Park, CA 94025, USA}
\author{Alexander F. Kemper}
%\affiliation{Lawrence Berkeley National Lab, 1 Cyclotron Road, Berkeley, CA 94720, USA}
\author{Brian Moritz}
%\affiliation{Stanford Institute for Materials and Energy Science (SIMES),
%SLAC National Accelerator Laboratory, Menlo Park, CA 94025, USA}
%\affiliation{Department of Physics and Astrophysics, University of North Dakota, Grand Forks, ND 58202, USA}
%\affiliation{Department of Physics, Northern Illinois University, DeKalb, IL 60115, USA}
\author{James K. Freericks}
%\affiliation{Department of Physics, Georgetown University, Washington, DC 20057, USA}
\author{Zhi-Xun Shen}
%\affiliation{Stanford Institute for Materials and Energy Science (SIMES),
%SLAC National Accelerator Laboratory, Menlo Park, CA 94025, USA}
%\affiliation{Departments of Physics and Applied Physics,
%Stanford University, Stanford, California 94305, USA}
%\affiliation{Geballe Laboratory for Advanced Materials, Stanford University, Stanford, CA 94305}
\author{Thomas P. Devereaux}
%\affiliation{Stanford Institute for Materials and Energy Science (SIMES),
%SLAC National Accelerator Laboratory, Menlo Park, CA 94025, USA}
%\affiliation{Geballe Laboratory for Advanced Materials, Stanford University, Stanford, CA 94305}

\title{\revision{Examining electron-boson coupling using time-resolved spectroscopy}: Supplementary Material}
\maketitle

\setcounter{section}{0}
\renewcommand{\thesection}{S\arabic{section}}
\setcounter{figure}{0}
\renewcommand{\thefigure}{S\arabic{figure}}
\setcounter{equation}{0}
\renewcommand{\theequation}{S\arabic{equation}}

\section{Influence of probe pulse width}

\begin{figure}[h!t]
        \includegraphics[width=0.8\columnwidth]{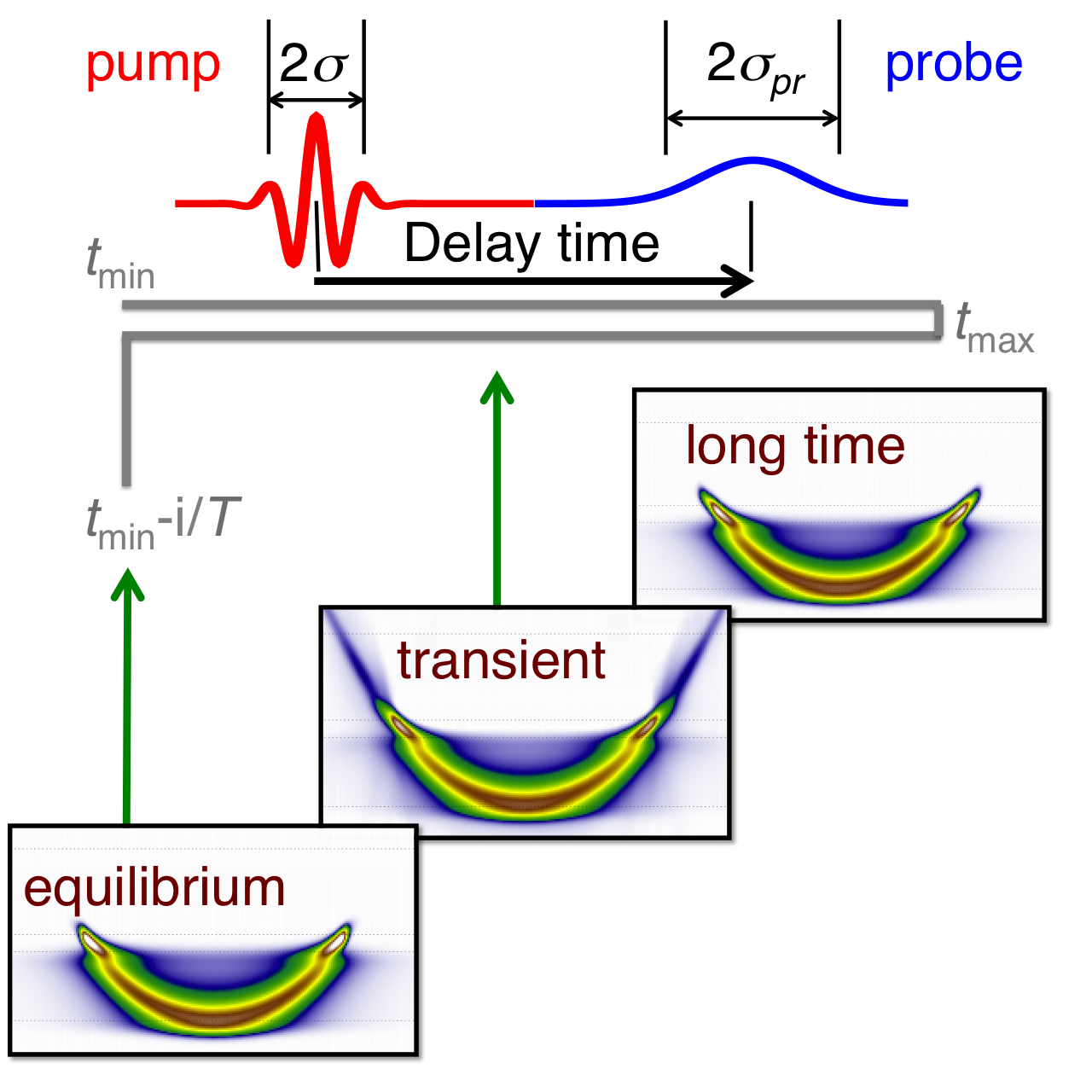}
        \caption{{\bf Pump-probe photoemission setup with Keldysh contour.} The time-resolved pump-probe photoemission is modeled using nonequilibrium Green functions defined on the Keldysh contour. The contour runs from the initial time $t_{\text{min}}$ to a maximum time $t_{\text{max}}$ covering the time span of the pump pulse (red) of temporal width $\sigma$ and the probe pulse (blue, only Gaussian envelope shown here) of width $\sigma_{pr}$, then back to $t_{\text{min}}$, and from there along the imaginary time direction to $t_{\text{min}}-\text{i}/T$. The tr-ARPES snapshots show the electronic structure in equilibrium, in the transient regime at short delay times, and in the long-time limit. The red pump pulse profile shows the actual few-cycle pulse used in the calculations.
        }
        \label{fig1}
\end{figure}

\begin{figure*}[h!t]
        \includegraphics[width=\textwidth]{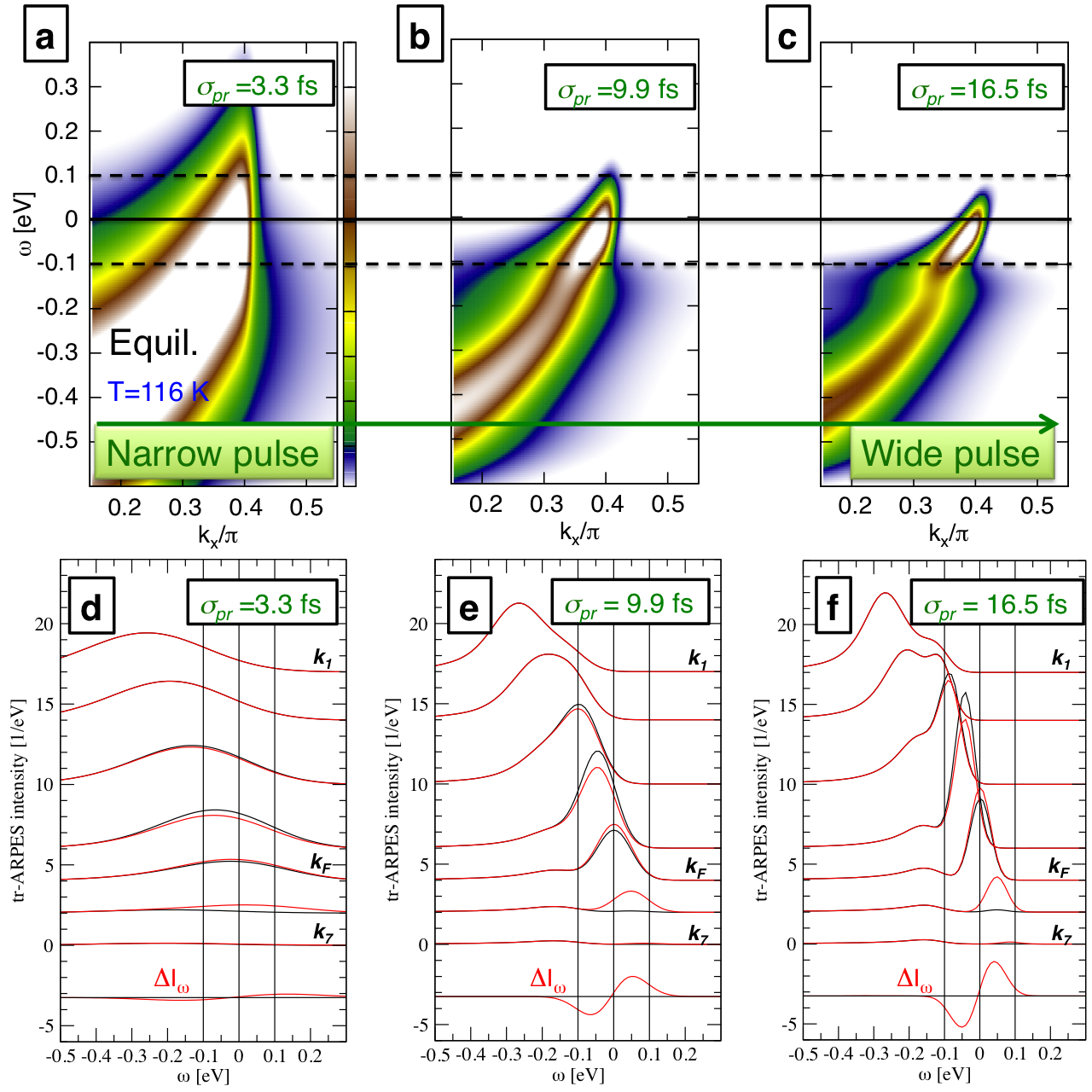}
        \caption{{\bf Sharpening around phonon edges for increasing energy resolution.}
        (a), (b), (c) Equilibrium ARPES momentum cuts (same as in Figure 1 of the main text) for different probe pulse resolutions as indicated.
        The tr-ARPES signals become sharper for broader probe pulses.
        (d), (e), (f) Corresponding energy distribution curves for selected momenta (same as in Figure 1).
        The restriction of slowly decaying spectral regions to the energy range inside the optical phonon
        range becomes more apparent when the temporal probe pulse width is increased, stressing the importance of the balance between time and energy resolution in tr-ARPES.
        }
        \label{figsupp1}
\end{figure*}

For the interpretation of time-resolved photoemission experiments, it is of central importance to include the influence of $\sigma_{pr}$, the temporal width of the probe pulse,\cite{Freericks09TRARPES} which is illustrated in the pump-probe setup in Figure \ref{fig1}. A narrower probe pulse implies a better time resolution but a worse energy resolution, and vice versa. This is illustrated in Figure \ref{figsupp1}, which shows the probe pulse width dependence of tr-ARPES intensity maps and corresponding EDCs. Apparently, the necessary energy resolution to resolve the phonon kink and the sharpening of spectral changes to be inside the phonon range has to be achieved by trading off intrinsic time resolution. 

\section{Breakdown of effective temperature description}
\begin{figure}[h!t]
        \includegraphics[width=\columnwidth]{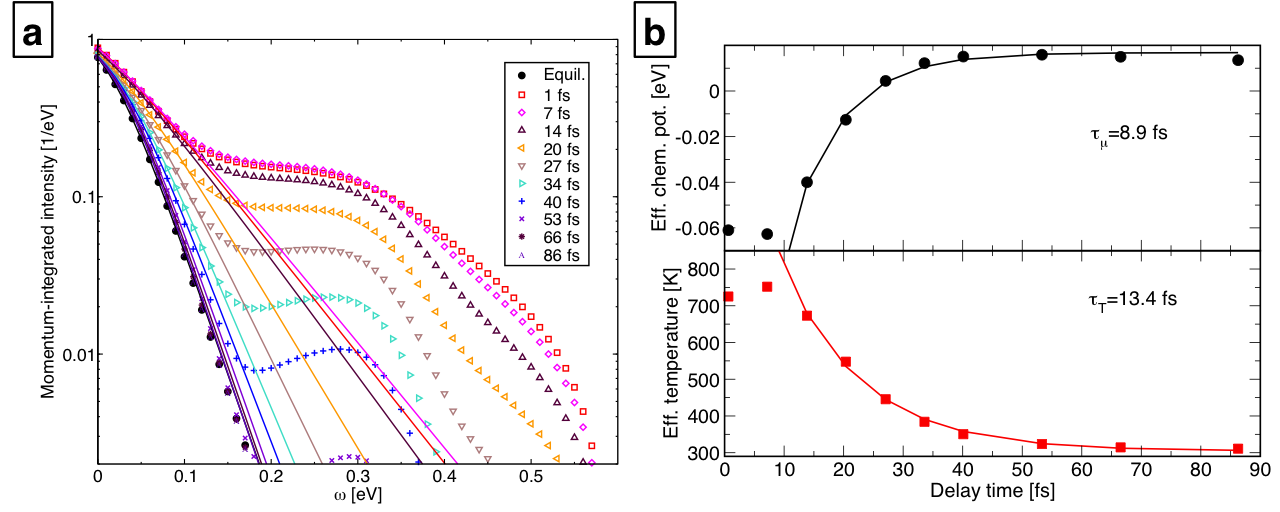}
        \caption{{\bf Effective temperatures.}
        (a) Momentum-integrated tr-ARPES data for weak pump (see Figure 2 of main text) with fits to Fermi functions $A/(\exp((\omega-\mu_{\text{eff}}(t))/(k_B T_{\text{eff}}(t)))+1)$ for time-dependent effective temperatures $T_{\text{eff}}(t)$ and chemical potentials $\mu_{\text{eff}}(t)$ in the energy window between 0 and 0.1 eV. (b) Extracted $\mu_{\text{eff}}(t)$ and $T_{\text{eff}}(t)$ as a function of delay time. The solid lines are fits to decaying exponentials starting at a delay time of 13 fs with decay constants as indicated.
        }
        \label{figeff}
\end{figure}

Figure \ref{figeff} shows fits of Fermi functions with effective temperatures and chemical potentials to the low-energy structure of the tr-ARPES energy distribution curves. This figure serves to show that one \emph{can} make such a fit in a small energy window and extract decay constants for the time-dependent temperatures and chemical potentials. However, these do not provide an adequate description of the excited system, which is not at thermal equilibrium. Notably, the time-dependent spectra deviate considerably from the fits already at slightly higher binding energies than the fit window. The figure therefore demonstrates the expected breakdown of an effective temperature description and calls for a more advanced data analysis. Additionally, we note that even if a Fermi function fit seems to work (see, e.g., Ref.\ \onlinecite{Crepaldi2012}), it is easily shown that the extracted decay time for the effective temperature is not related in a simple fashion to energy-dependent decay times for the spectral intensities. This is due to the fact that the intensity at a given binding energy depends nonlinearly on the effective temperature in this case, namely through a Fermi function. Therefore it is doubtful which information can be gained from the temperature decay time.

\section{Relaxation rates}
\begin{figure}[h!t]
        \includegraphics[width=0.7\columnwidth]{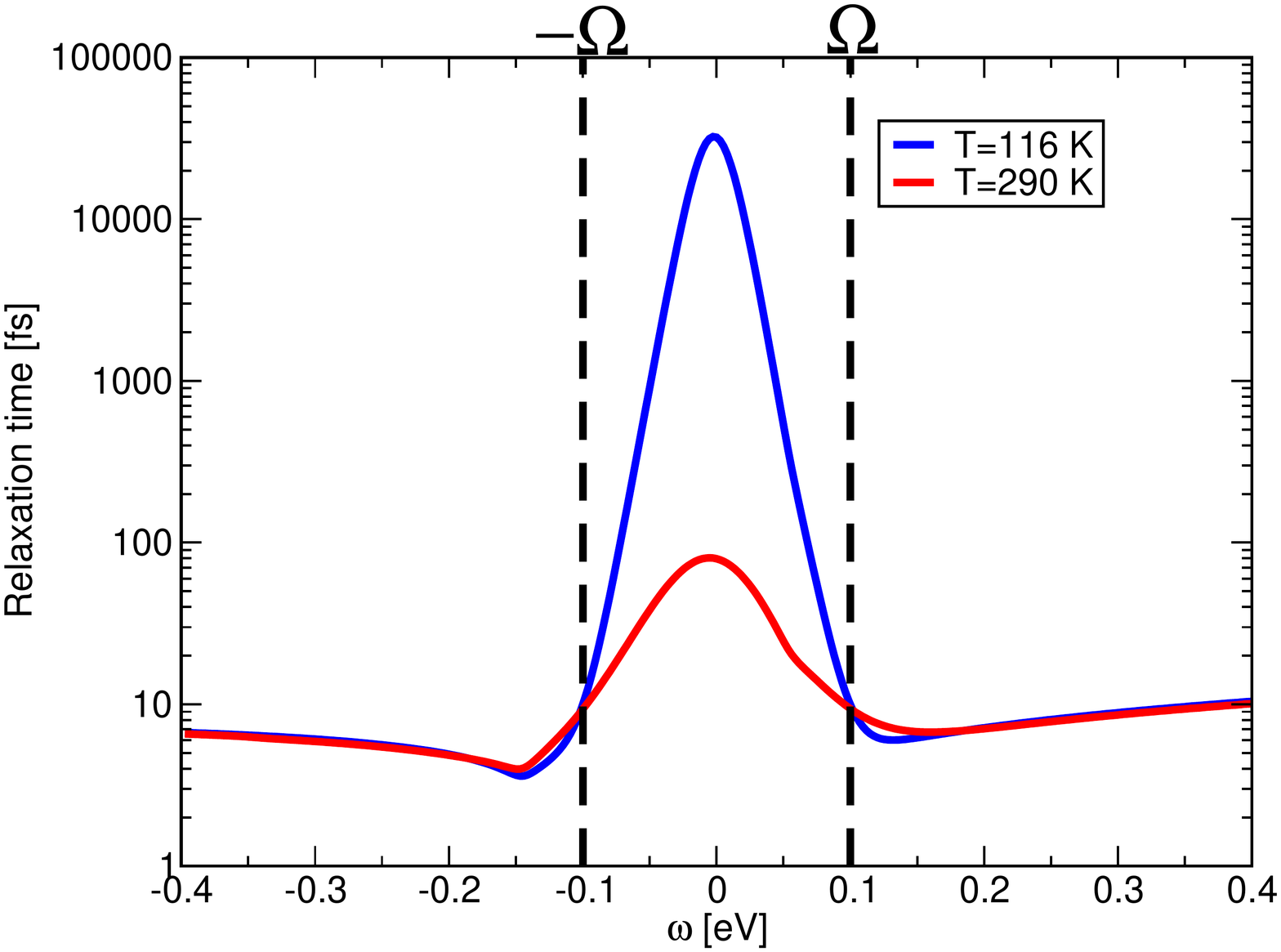}
        \caption{{\bf Equilibrium temperature dependent relaxation times.}
        The relaxation times given by the self-energy (without additional frequency broadening) for the two equilibrium temperatures (116 K and 290 K) used in this paper. 
        }
        \label{figsupp2}
\end{figure}

In Figure 1 of the main text, we showed how the relaxation process inside the phonon window is strongly temperature dependent. For low temperatures, the time scales become considerably longer and thus lead to a slower relaxation. Figure \ref{figsupp2} shows the corresponding equilibrium relaxation times. Obviously, the times due to electron-phonon scattering are basically independent of temperature outside the phonon window. Inside the phonon window, however, they are strongly temperature dependent. The time scales for our model parameters are well below a picosecond for the higher equilibrium temperature $T$ $=$290 K, but become as large as tens of picoseconds for $T$ $=$ 116 K.

The relaxation times shown in Figure \ref{figsupp2} are obtained from the imaginary part of the equilibrium retarded self-energy according to
\begin{align}
\tau(\omega) = \frac{-\hbar}{2 \; \text{Im} \; \Sigma^R(\omega)},
\label{eq:relax}
\end{align}
where
\begin{align}
\im \Sigma^R(\omega) = -\pi g^2 \big[ & N(\omega + \Omega) \left( n_B(\Omega) + n_F(\Omega+\omega) \right) \notag \\
				+ & N(\omega - \Omega) \left( n_B(\Omega) + n_F(\Omega-\omega) \right ) \big],
\label{eq:sigma}
\end{align}
and $N(\omega)$ $=$ $L^{-1}\sum_{\kk} \delta(\omega-\epsilon(\kk))$ (the sum is over the two-dimensional Brillouin zone, with $L$ the number of momentum points summed over) is the bare electronic density of states.\cite{Cuk2005} The relaxation rate $\Gamma(\omega)$ $=$ $1/\tau(\omega)$ is thus of the form $g^2 \phi(\omega)$, which is used for the extraction of $g^2$ from the tr-ARPES data in the main text. Here, the equilibrium self-energy is taken at the initial temperature of the pump-probe simulation.

For the relaxation rates shown in Figure 4 of the main text, we apply an additional averaging procedure to account for the energy resolution due to the finite width of the probe pulse. The averaged self-energy is obtained by convoluting its inverse with a resolution function:
\begin{align}
\frac{1}{\im \Sigma_{\text{aver}}(\omega)} = \int \text{d}\omega' \frac{1}{\sigma_{\text{res}}\sqrt{2\pi}} \; exp\left(\frac{-\omega'^2}{2\sigma_{\text{res}}^2}\right) \frac{1}{\im \Sigma^R(\omega')}.
\end{align}
For the probe pulse width $\sigma_{pr}$ $=$ 16.5 fs, the corresponding energy resolution is given by $\sigma_{\text{res}}$ $=$ $\sqrt{2}\hbar/\sigma_{pr}$ $\approx$ 57 meV. Note that this is in contrast to Matthiessen's rule for scattering times, which states that inverses of times should be averaged instead of the times themselves. The difference is that Matthiessen's rule applies for averages of scattering processes for various sources of scattering, all for the same particle. Here, however, we are interested in averages of time scales for different particles. The reason why one should average the relaxation times here, instead of the rates, is the following: In the long-time limit, the slowest decaying components determine the effective relaxation rate. To give those components with the longest time scales the largest weight, one has to average times, not rates. The same averaging procedure is applied to yield $\phi_{\text{aver}}$, which is used to extract $g^2$ in Figure 4b of the main text.

\begin{figure}[h!t]
        \includegraphics[width=0.7\columnwidth]{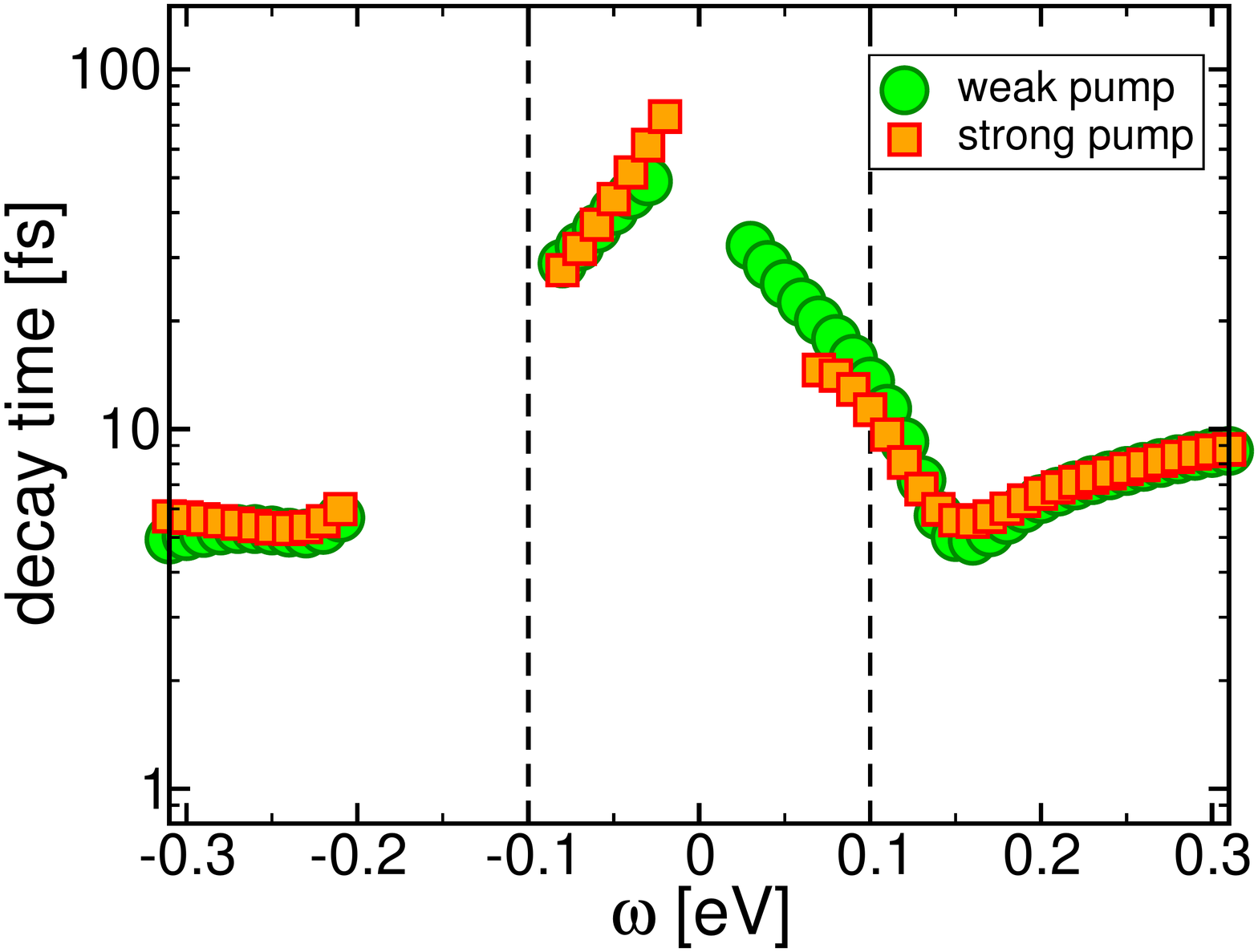}
        \caption{{\bf Comparison of tr-ARPES relaxation times for weak and strong pumps.}
        For the weak and strong pump fields (same as in Figure 2 of main text), the energy dependence of tr-ARPES relaxation times is shown here. 
        }
        \label{figsupp4}
\end{figure}

Finally, we demonstrate that the relaxation times for particular energies do not significantly depend on the pump field strength as long as one does not measure times too close to the application of the pump. To this end, we show in Figure \ref{figsupp4} a comparison of the time scales for exponential fits to the time traces of the energy-dependent tr-ARPES intensities for the weak and strong pumps (see Figure 2 of main text). The times extracted for weak and strong pump pulses agree reasonably well. This overall agreement of the time scales demonstrates that the interpretation of the nonequilibrium dynamics in terms of equilibrium electron-phonon scattering rates is valid even beyond the weak pump regime.

\section{Inclusion of electron-electron scattering}

Here we present the effects of an additional electron-electron scattering channel by adding a second-order perturbative local Coulomb repulsion of strength $U$ in the local approximation to the electron-phonon self-energy:
\begin{align}
\Sigma^U(t,t') = U^2 G^0(t,t') G^0(t,t') G^0(t',t),
\end{align}
where $G^0$ $=$ $\sum_\kk G^0_\kk$ is the bare local electronic Green function. Note that this representation is equivalent to summing the full momentum-dependent $\Sigma^U_\kk(t,t')$ over momenta. As Coulomb repulsion values representative of weak to moderate coupling we choose $U$ = 0.25 eV and 0.5 eV for our calculations, which correspond to $W/8$ and $W/4$, where $W$ is the bare bandwidth of the electrons.

\begin{figure}[h!t]
        \includegraphics[width=0.7\columnwidth]{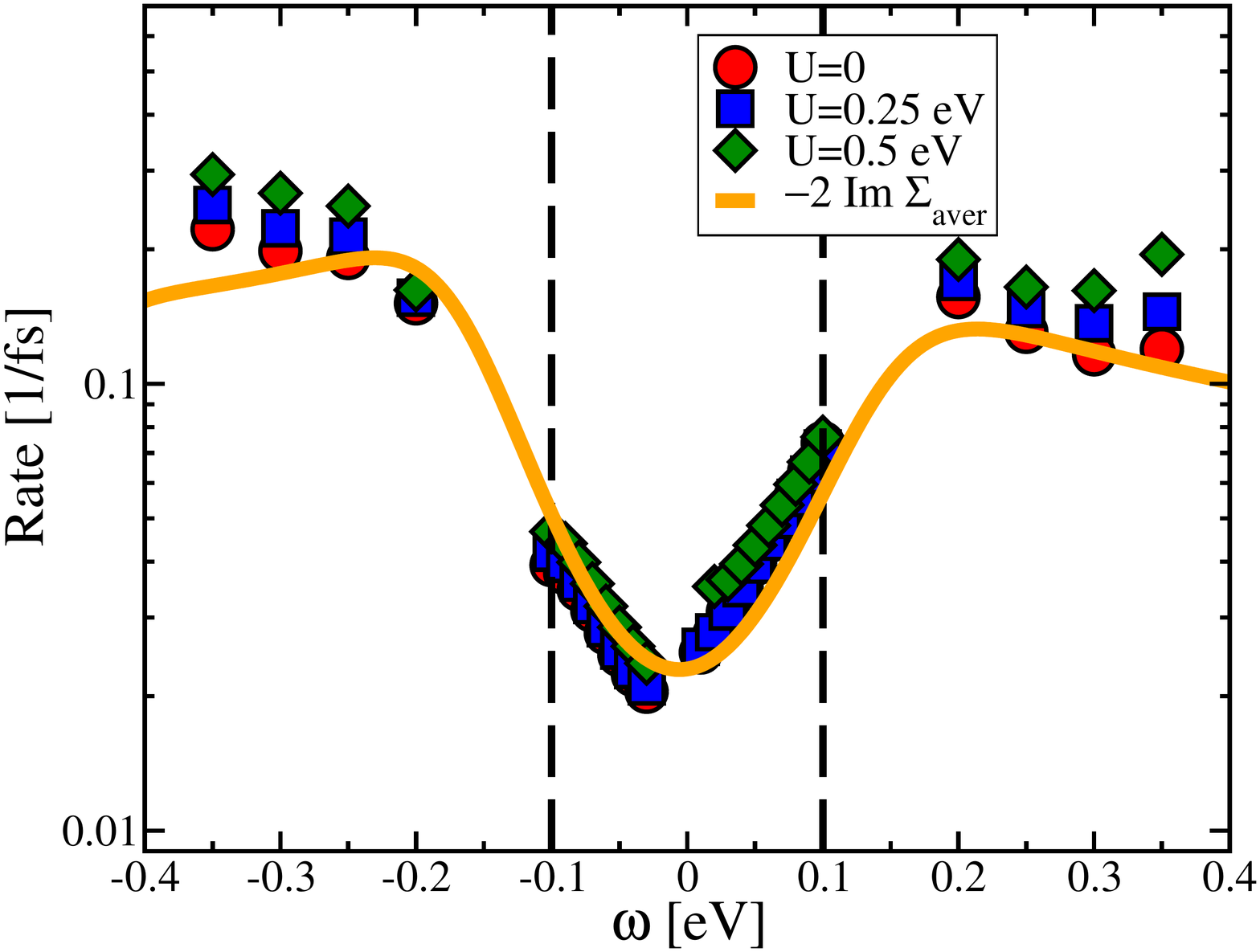}
        \caption{{\bf Comparison of tr-ARPES relaxation times without and with Coulomb repulsion.}
        tr-ARPES relaxation rates including 2nd order electron-electron interaction diagram for local Coulomb repulsion $U$ as indicated. The rates do increase, but the phonon window effect is still clearly visible. 
        }
        \label{figsupp6}
\end{figure}

\begin{figure}[h!t]
        \includegraphics[width=0.7\columnwidth]{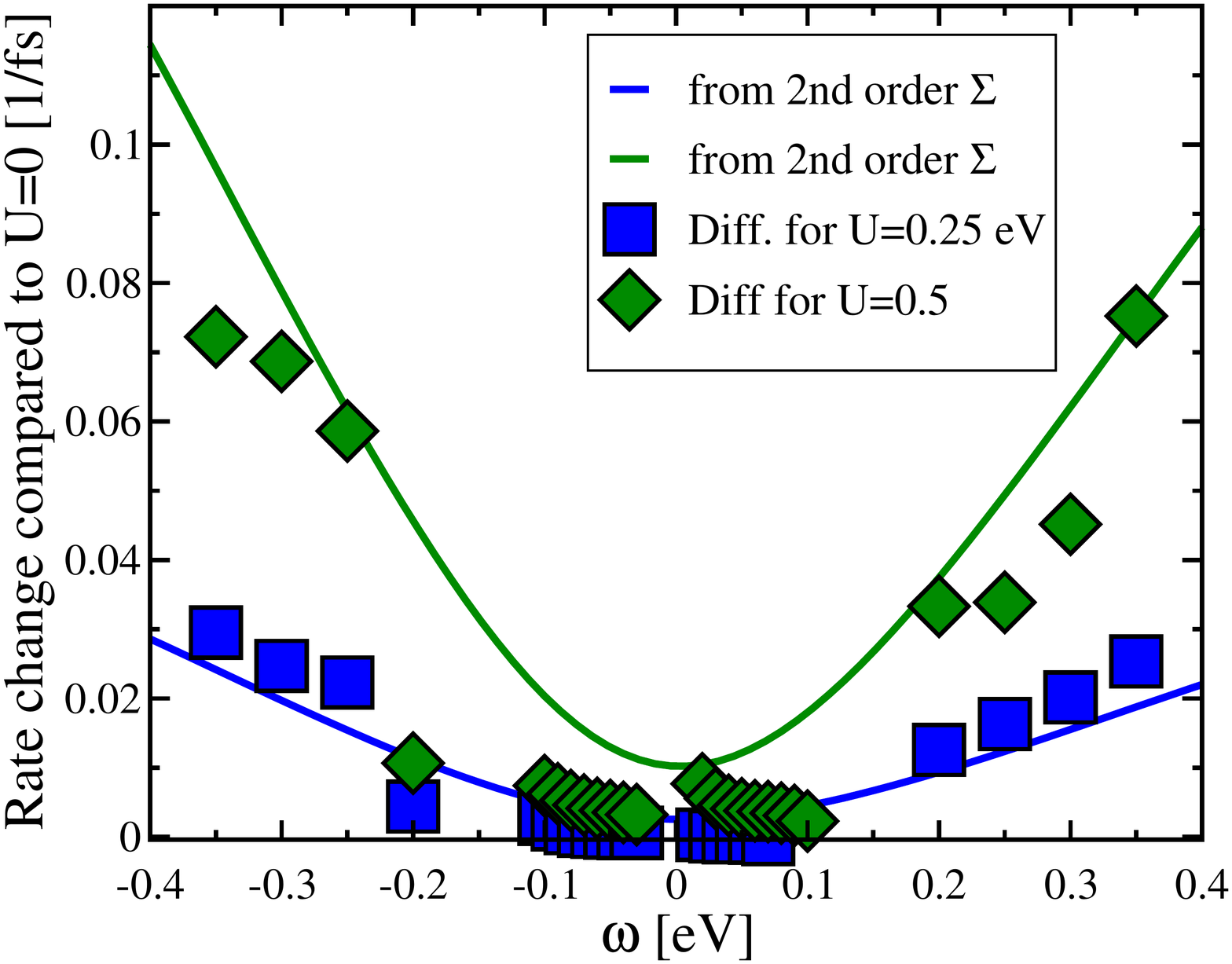}
        \caption{{\bf Additive nature of electron-phonon and electron-electron scattering channels.}
        Difference between $U$ $>$ 0 and $U$ $=$ 0 relaxation rates from Figure \ref{figsupp6} (symbols), compared with 2nd order electron-electron equilibrium scattering rates (solid lines), showing that rates for different scattering channels are additive (Matthiessen's rule). 
        }
        \label{figsupp7}
\end{figure}

\begin{figure}[h!t]
        \includegraphics[width=0.7\columnwidth]{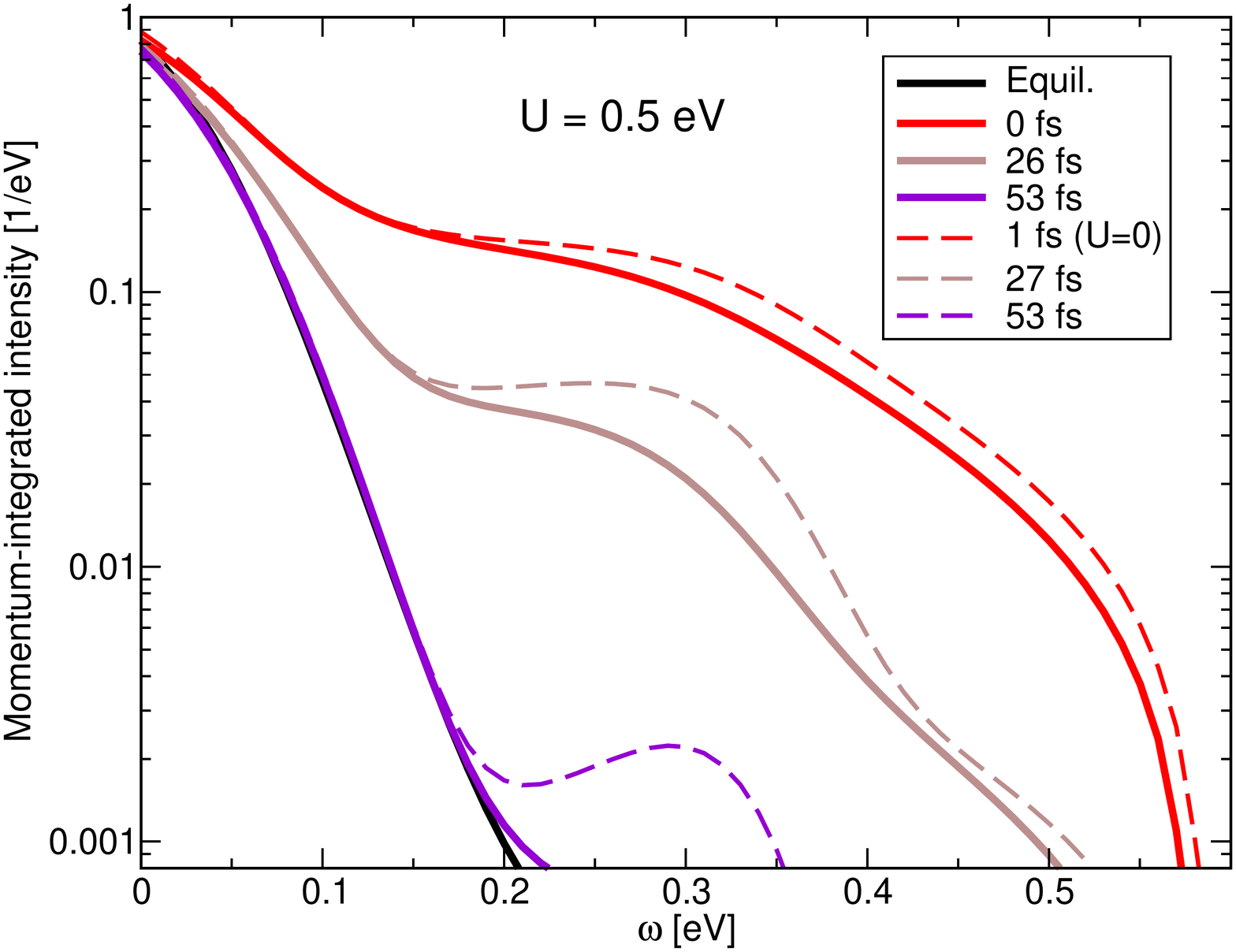}
        \caption{{\bf Transient EDCs in the presence of Coulomb repulsion.}
        Transient EDCs in the presence of $U$ $=$ 0.5eV Coulomb repulsion show better agreement with a higher temperature Fermi-Dirac distribution than for pure electron-phonon case (dashed lines, reproduced from Figure \ref{figeff}), still show a (smaller) nonthermal feature even when the pump and probe pulses do not overlap (53 fs). 
        }
        \label{figsupp8}
\end{figure}

The analysis of these results shows that a treatment of electron-electron scattering on the same level as we treat electron-phonon scattering supports our main results, namely the phonon window effect for slow versus fast relaxation (Figure \ref{figsupp6}) and the long-time relaxation being governed by equilibrium scattering rates. Even though the inclusion of electron-electron scattering does alter the short-time transient behavior and spectral redistribution (see Figure \ref{figsupp8}) which are not part of the of the relaxation analysis from the outset, it does not alter our main conclusions regarding the relaxation back to equilibrium. This follows from the additive nature of the individual relaxation channels (fulfilling Matthiessen's rule). The additive nature of the long-time relaxation rates is demonstrated in Figure \ref{figsupp7} by subtracting the relaxation rates of electron-phonon scattering alone from the rates where both scattering channels are present, and comparing them with the equilibrium scattering rates for electron-electron scattering. We also note that even going beyond this second order approximation for the electron-electron scattering, the importance of electron-phonon coupling for thermal relaxation would not diminish. A fully energy-conserving treatment of electron-electron scattering precludes full relaxation by electron-electron scattering alone, highlighting the importance of electron-phonon coupling for relaxation in pump-probe spectroscopies on solids.

\bibliography{bib}{}

%%%%%%%%%%%%%%%%%%%%%%%%%%%%%%%%%%%%%%%%%%%%%%%%%%%%%%%
%%%%% FIGURES     OF        SUPPLEMENT     %%%%%%%%%%%%%%%%%%%%%%%%%%%%%%%
%%%%%%%%%%%%%%%%%%%%%%%%%%%%%%%%%%%%%%%%%%%%%%%%%%%%%%%
%
%
\newpage
\newpage
\newpage
\newpage
\newpage
\newpage
\newpage
\newpage
